\documentclass[reprint,prl,twocolumn,superscriptaddress,showpacs,showkeys,floatfix,preprintnumbers]{revtex4-1}
\usepackage{braket}
\usepackage{xargs}
\usepackage{mathtools}
\usepackage[english]{babel}
\usepackage[utf8]{inputenc}
\usepackage{amsmath,amssymb,braket,slashed,mathrsfs}
\usepackage{pifont}
\usepackage{MnSymbol}
\usepackage{graphicx}
\graphicspath{{./fig/}}
\usepackage{tikz}
\usetikzlibrary{shapes,arrows,positioning}
\tikzset{decision/.style={diamond, draw, fill=blue!20, text width=4.5em, text badly centered, inner sep=0pt}}
\tikzset{block/.style={rectangle, draw, fill=blue!20, text width=10em, text centered, rounded corners, minimum width=3.5cm}}
\tikzset{block1/.style={rectangle, draw, fill=blue!20, text width=18.5em, text centered, rounded corners, minimum width=3.5cm}}
\tikzset{line/.style={draw, -latex, thick}}
\usepackage{color,hyperref}
\usepackage{multirow,array}

\usepackage{cleveref}

\newcommand{\innovation}{Collaborative Innovation Center of Quantum Matter, Beijing 100871, China}
\newcommand{\chep}{Center for High Energy Physics, Peking University, Beijing 100871, China}
\newcommand{\pkuphy}{School of Physics, Peking University, Beijing 100871,
China}

\newcommand{\Uconn}{Department of Physics, University of Connecticut, Storrs, CT 06269, USA}
\newcommand{\RBRC}{RIKEN-BNL Research Center, Brookhaven National Laboratory, Building 510, Upton, NY 11973} 
\newcommand{\KPH}{Institut f\"ur Kernphysik, Johannes Gutenberg-Universit\"{a}t,
	J.J. Becher-Weg 45, 55128 Mainz, Germany}
\newcommand{\PRISMA}{PRISMA$+$ Cluster of Excellence, Johannes Gutenberg-Universit\"{a}t, Mainz, Germany}
\newcommand{\FRIB}{Facility for Rare Isotope Beams, Michigan State University, East Lansing, MI 48824, USA}
\newcommand{\UW}{Department of Physics, University of Washington,
	Seattle, WA 98195-1560, USA}
\newcommand{\UK}{Department of Physics and Astronomy, University of Kentucky, Lexington, Kentucky 40506, USA}
\newcommand{\LBNL}{Nuclear Science Division, Lawrence Berkeley National Laboratory, Berkeley, CA 94720, USA}

\begin{document}
\title{Lattice QCD Calculation of Electroweak Box Contributions to Superallowed Nuclear and Neutron Beta Decays}

\author{Peng-Xiang~Ma}\affiliation{\pkuphy}
\author{Xu~Feng}\email{xu.feng@pku.edu.cn}\affiliation{\pkuphy}\affiliation{\innovation}\affiliation{\chep}
\author{Mikhail~Gorchtein}\affiliation{\KPH}\affiliation{\PRISMA}
\author{Lu-Chang~Jin}\email{ljin.luchang@gmail.com}\affiliation{\Uconn}\affiliation{\RBRC}
\author{Keh-Fei~Liu}\affiliation{\UK}\affiliation{\LBNL}
\author{Chien-Yeah~Seng}\affiliation{\FRIB}\affiliation{\UW}
\author{Bi-Geng~Wang}\affiliation{\UK}\affiliation{\LBNL}
\author{Zhao-Long~Zhang}\affiliation{\pkuphy}
%\pacs{PACS}
%

\date{\today}

\begin{abstract}
We present the first lattice QCD calculation of the universal axial $\gamma W$-box contribution $\Box_{\gamma W}^{VA}$ to both superallowed nuclear and neutron beta decays. 
This contribution emerges as a significant component within the theoretical uncertainties surrounding the extraction of $|V_{ud}|$ from superallowed decays.
Our calculation is conducted using two domain wall fermion ensembles at the physical pion mass. 
To construct the nucleon 4-point correlation functions, we employ the random sparsening field technique. Furthermore, we incorporate long-distance contributions to the hadronic function using the infinite-volume reconstruction method.
Upon performing the continuum extrapolation, we arrive at $\Box_{\gamma W}^{VA}=3.65(7)_{\mathrm{lat}}(1)_{\mathrm{PT}}\times10^{-3}$. Consequently, this yields a slightly higher value of $|V_{ud}|=0.97386(11)_{\mathrm{exp.}}(9)_{\mathrm{RC}}(27)_{\mathrm{NS}}$, 
reducing the previous $2.1\sigma$ tension with the CKM unitarity to $1.8\sigma$.
Additionally, we calculate the vector $\gamma W$-box contribution to the axial charge $g_A$, denoted as $\Box_{\gamma W}^{VV}$, and explore its potential implications.
\end{abstract}

\maketitle

\noindent
{\bf Introduction}: The Cabibbo-Kobayashi-Maskawa (CKM) matrix plays a crucial role as a fundamental ingredient of the standard model (SM) where its unitarity is expected. 
Recent findings indicate a $2.1\sigma$ tension in the examination of the first row's unitarity~\cite{Workman:2022ynf}
\begin{equation}
|V_{ud}|^2 + |V_{us}|^2 + |V_{ub}|^2 = 0.9985(6)_{|V_{ud}|}(4)_{|V_{us}|}.
\end{equation}
Depending on the choice of inputs, tensions can be $3\sigma$ or larger~\cite{FlavourLatticeAveragingGroupFLAG:2021npn}.
Further efforts to accurately determine the CKM matrix elements could unveil new signals and deepen our understanding of the underlying physics.

\begin{figure}[htb]
\centering
\includegraphics[width=0.35\textwidth,angle=0]{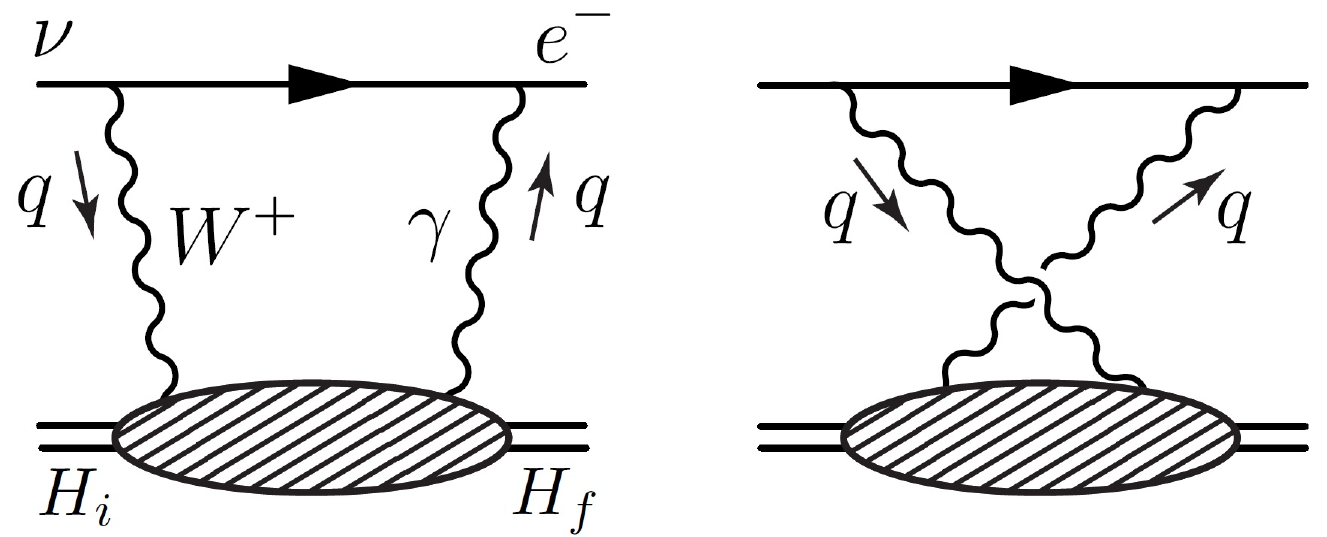}
\caption{The $\gamma W$-box
diagrams for the semileptonic decay process $H_i\to H_fe\bar{\nu}_e$.}
\label{fig:photon_W_diags}
\end{figure}

The top-left corner element $|V_{ud}|$ is most precisely extracted from neutron and superallowed nuclear beta decays. The latter currently claims the highest experimental accuracy, but the former is catching up thanks to improvements in experimental precision~\cite{UCNA:2017obv,Markisch:2018ndu,Hassan:2020hrj,UCNt:2021pcg}.
On the theory side, superallowed nuclear decays suffer from extra nuclear-structure-dependent uncertainties that are recently under careful  scrutiny~\cite{Seng:2018yzq,Gorchtein:2018fxl,Seng:2022cnq}, while neutron decay is theoretically cleaner. 
Nevertheless, a major source of uncertainty common to both cases is the single-nucleon $\gamma W$-box diagrams (see Fig.~\ref{fig:photon_W_diags}) that renormalize the neutron vector and axial charges~\cite{Sirlin:1977sv}. They take the following forms~\cite{Gorchtein:2021fce}~\footnote{In Ref~\cite{Gorchtein:2021fce}, the superscripts ``V'' and ``A'' in $\Box_{\gamma W}^V$ and $\Box_{\gamma W}^A$ are used to denote the correction to the coupling $g_V$ and $g_A$, respectively. Meanwhile, in this paper the second superscript in
 $\Box_{\gamma W}^{VA}$ and $\Box_{\gamma W}^{VV}$ is used to denote whether the axial or vector component of the charged weak current is taken.
The matching of the two notations is: $\Box_{\gamma W}^V=\Box_{\gamma W}^{VA}$, $\Box_{\gamma W}^A=\Box_{\gamma W}^{VV}$.}:
\begin{equation}
\begin{aligned}
&\Box_{\gamma W}^{VA}=\frac{ie^2}{2m_N^2}\int \frac{d^4 q}{(2\pi)^4}\frac{m_W^2}{m_W^2-q^2}\frac{\epsilon^{\mu\nu\alpha\lambda}q_{\alpha}p_{\lambda}}{(q^2)^2}T_{\mu\nu}^{\gamma W}, \\
&\Box_{\gamma W}^{VV}=-\frac{ie^2}{2m_N^2{\mathring g}_A}\int \frac{d^4 q}{(2\pi)^4}\frac{m_W^2}{m_W^2-q^2}\frac{\epsilon^{\mu\nu\alpha\lambda}q_{\alpha}S_{\lambda}}{(q^2)^2}T_{\mu\nu}^{\gamma W},\\
&T^{\gamma W}_{\mu\nu}=\int d^4x e^{iq\cdot x}\bra{p(p,S)}T\left\{J^{em}_{\mu}(x)J^{W}_{\nu}(0)\right\}\ket{n(p,S)},
\end{aligned}
\end{equation}  
where $p$ and $S$ are the momentum and spin vectors of the nucleon states.
${\mathring g}_A$ is the nucleon axial charge, where the symbol $\mathring{}$ indicates that it is defined in the isospin limit.
$J^{em}_{\mu}$ is the electromagnetic current and $J^{W}_{\mu}$ the weak current with
$J^{W,V}_{\mu}$ and $J_{\mu}^{W,A}$ its vector and axial-vector part, respectively. 
Specifically, in $0^{+}\to 0^{+}$ superallowed beta decay, only $\Box_{\gamma W}^{VA}$ comes into play, while in $\frac{1}{2}^{+}\to \frac{1}{2}^{+}$ free neutron beta decay, both $\Box_{\gamma W}^{VA}$ and $\Box_{\gamma W}^{VV}$ would contribute. 

Various approaches to computing $\Box_{\gamma W}^{V A}$ range from earlier works by Marciano and Sirlin~\cite{Marciano:1985pd,Marciano:2005ec}
to more recent dispersive analyses by Seng et al~\cite{Seng:2018yzq,Seng:2018qru}.
The latter, in particular, improved the nonperturbative contribution for loop momentum square $Q^2\le2$ GeV$^2$ and unveiled the
tension with the first-row CKM unitarity, which was also observed in several follow-up works~\cite{Czarnecki:2019mwq,Seng:2020wjq,Shiells:2020fqp,Hayen:2020cxh}. 
In the meantime, while the radiative correction to the axial charge $g_A$ does not directly affect the $|V_{ud}|$ extraction, it is necessary for
comparing the experimentally measured $g_A$ with that computed using lattice QCD. 
The study of $\Box_{\gamma W}^{VV}$ 
has so far included estimations inspired by the holographic QCD model~\cite{Hayen:2020cxh} and dispersion relations~\cite{Gorchtein:2021fce}.

Lattice QCD offers a direct nonperturbative approach to compute the box correction $\Box_{\gamma W}^{VA}$, especially for $Q^2\le2$ GeV$^2$. First lattice calculations of $\Box_{\gamma W}^{VA}$ were successfully conducted in the pion~\cite{Feng:2020zdc} and kaon channel~\cite{Seng:2020jtz,Ma:2021azh}, and have recently been confirmed by an independent lattice calculation~\cite{Yoo:2023gln}. The data reported in~\cite{Feng:2020zdc} were also used for a joint lattice QCD - dispersion relation analysis~\cite{Seng:2020wjq}. This letter extends this calculation to the neutron decay channel, which entails a direct computation of the nucleon four-point function at the physical pion mass. We also briefly discuss our numerical result of $\Box_{\gamma W}^{VV}$, and its implication to the radiative correction to axial charge.\\

\noindent
{\bf Methodology}: The notations used in this work align with those used in \cite{Feng:2020zdc}.
We define the hadronic function $H^{VA}_{\mu \nu}$ within Euclidean space
\begin{equation}
\mathcal{H}^{VA}_{\mu\nu}(t,\vec{x})\equiv
\langle H_f|T\left[J^{em}_\mu(t,\vec{x})J^{W,A}_\nu(0)\right]|H_i\rangle,
\end{equation}
where $H_{i/f}$ represents the zero-momentum projected neutron/proton state, created by a smeared-source nucleon operator.
The computation of box contribution $\Box_{\gamma
W}^{VA}$ involves a momentum integral
\begin{equation}
\label{eq:master}
\Box_{\gamma
W}^{VA}=\frac{3\alpha_e}{2\pi}\int\frac{dQ^2}{Q^2}\,\frac{m_W^2}{m_W^2+Q^2}M_n(Q^2).
\end{equation}
$M_n(Q^2)$ is a weighted integral of the hadronic function $H(t,\vec{x})=\epsilon_{\mu\nu\alpha0}x_\alpha \mathcal{H}^{VA}_{\mu \nu}(t,\vec{x})$, defined as
\begin{equation}
\label{eq:integral}
M_n(Q^2)=-\frac{1}{6}\frac{\sqrt{Q^2}}{m_N}\int d^4x\,\omega(t,\vec{x})H(t,\vec{x}),
\end{equation}
with $m_W$ and $m_N$ the masses of the $W$-boson and the nucleon. The weighting function is
\begin{equation}
\omega(t,\vec{x})=\int_{-\frac{\pi}{2}}^{\frac{\pi}{2}}\frac{\cos^3\theta\,d\theta}{\pi}\frac{j_1\left(|\vec{Q}||\vec{x}|\right)}{|\vec{x}|}
\cos\left(Q_0 t \right),
\end{equation}
where $|\vec{Q}|=\sqrt{Q^2} \cos\theta$, $Q_0=\sqrt{Q^2}\sin\theta$ and $j_n(x)$ are the spherical Bessel functions. 

To evaluate $M_n(Q^2)$ as prescribed in Eq.~(\ref{eq:integral}), it is necessary to extend the temporal integration range sufficiently to reduce truncation effects. However, as the time separation between the two currents increases, the lattice data tend to exhibit greater noise-to-signal ratio. Here we employ the infinite volume reconstruction (IVR) method~\cite{Feng:2018qpx} to incorporate the long-distance (LD) contribution arising from the region where $|t|>t_s$. Here, 
$t_s$ is the time slice at which the short-distance (SD) and LD contributions are separated.
The IVR method, in addition to eliminating the power-law suppressed finite volume error, can also reduce the lattice statistical error in the long distance region.
To elaborate, we divide the integral into SD contribution, weighted by $\omega(t,\vec{x})$, and LD contribution, weighted by $\tilde{\omega}(t,\vec{x})$
\begin{equation}
\label{eq:long&short}
M_n(Q^2)=M_n^{\text{SD}}(Q^2,t_s)+M_n^{\text{LD}}(Q^2,t_s,t_g)
\end{equation}
with
\begin{align}
&M_n^{\text{SD}}(Q^2,t_s)=-\frac{1}{6}\frac{\sqrt{Q^2}}{m_N}\int_{-t_s}^{t_s} dt\int d^3\vec{x}\,\omega(t,\vec{x})H(t,\vec{x}),
\nonumber
\\
&M_n^{\text{LD}}(Q^2,t_s,t_g)=-\frac{1}{6}\frac{\sqrt{Q^2}}{m_N}\int d^3\vec{x}\,\tilde{\omega}(t_s,t_g,\vec{x})\bar{H}(t_g,\vec{x}),
\label{eq:MnLD_direct}
\end{align}
and
\begin{align}
\tilde{\omega}(t_s,t_g,\vec{x}) =& 2\int_{-\frac{\pi}{2}}^{\frac{\pi}{2}} \frac{\cos ^{3} \theta d \theta}{\pi} \frac{j_{1}\left(|\vec{Q}||\vec{x}| \right)}{|\vec{x}|}\times
\nonumber\\
& \operatorname{Re} \left(\frac{e^{-i Q_0 t_s}}{E_{\vec{Q}}-m_N+i Q_0}\right) e^{-(E_{\vec{Q}}-m_N)(t_s-t_g)}.
\end{align}
Here, $\bar{H}(t,\vec{x})=[H(t,\vec{x})+H(-t,\vec{x})]/2$, $E_{\vec{Q}}=\sqrt{m_N^2+|\vec{Q}|^2}$ and $t_g$ is chosen to be large enough to ensure the ground-intermediate-state dominance. Once $t_g$ is fixed, $t_s$ can be varied
to further verify the ground-state dominance.
In the final results, it is natural to choose $t_s=t_g$.

Due to the factor ${1}/{Q^2}$
in Eq.~(\ref{eq:master}), we observe that $\Box_{\gamma
W}^{VA}$ encounters a notably increased noise originating from $M_n(Q^2)$ at small $Q^2$ region. To mitigate this noise, 
we can use the model-independent relation
\begin{equation}
\label{eq:low_Q_relation}
\int d^3\vec{x}\,\bar{H}(t_g,\vec{x})=-3{\mathring g}_A({\mathring \mu}_p+{\mathring \mu}_n)
\end{equation}
to substitute $M_n^{\text{LD}}(Q^2,t_s,t_g)$ with
 \begin{align}
 \label{eq:MnLD_substitution}
M_n^{\text{LD}}
 &=-\frac{1}{6}\frac{\sqrt{Q^2}}{m_N}\int d^3\vec{x}\,\left[\tilde{\omega}(t_s,\vec{x})-\tilde{\omega}_0\right]\bar{H}(t_g,\vec{x})
 \nonumber\\
 &+\frac{1}{2}\frac{\sqrt{Q^2}}{m_N}\tilde{\omega}_0g_A(\mu_p+\mu_n).
 \end{align}
Above, as far as ground-state dominance is satisfied, $\bar{H}(t_g,\vec{x})$ is independent of $t_g$.
${\mathring \mu}_{p,n}$ denote the proton/neutron magnetic moments defined in the isospin limit. During
the substitution process, we incorporate experimentally measured values
for $g_A$ and $\mu_{p,n}$ as depicted in Eq.~(\ref{eq:MnLD_substitution}).
The difference is of a higher order and numerically negligible. Furthermore,
$\tilde{\omega}_0$ is defined as
\begin{eqnarray}
\tilde{\omega}_0&=&2\int_{-\frac{\pi}{2}}^{\frac{\pi}{2}} \frac{\cos ^{3} \theta d \theta}{\pi} \frac{|\vec{Q}|}{3}\operatorname{Re} \left(\frac{1}{\frac{Q^2}{2m_N}+i Q_0}\right)
\nonumber\\
&=&\frac{1}{3}\frac{2\sqrt{1+\tau}+\sqrt{\tau}}{(\sqrt{1+\tau}+\sqrt{\tau})^2},\quad\text{with $\tau=\frac{Q^2}{4m_N^2}$}.
\end{eqnarray}
Importantly, the convergence of the integral with $\tilde{\omega}
-\tilde{\omega}_0$ at $Q^2\to 0$ is considerably faster than that with $\tilde{\omega}$, enabling a more
efficient control over statistical uncertainties. 
We refer to the calculation of $M_n^{\text{LD}}$ using Eq.~(\ref{eq:MnLD_direct}) and Eq.~(\ref{eq:MnLD_substitution}) as the ``direct'' and ``substitution'' methods, respectively.

We introduce a four-momentum squared scale $Q^2_{\text{cut}}$ which separates the $Q^2$ integral into two regimes,
\begin{align}
\Box_{\gamma W}^{VA}&=\Box_{\gamma
W}^{VA,\le Q_{\mathrm{cut}}^2}+\Box_{\gamma W}^{VA,>Q_{\mathrm{cut}}^2}
\\
&=\frac{3\alpha_e}{2\pi}\left(\int_0^{Q^2_\mathrm{cut}}\frac{dQ^2}{Q^2}+\int_{Q^2_\mathrm{cut}}^\infty
\frac{dQ^2}{Q^2}\right)\frac{m_W^2}{m_W^2+Q^2}M_n(Q^2).
\nonumber
\end{align}
For $\Box_{\gamma
W}^{VA,\le Q_{\mathrm{cut}}^2}$ we use lattice results as inputs. Conversely, for $\Box_{\gamma
W}^{VA,> Q_{\mathrm{cut}}^2}$, we utilize the perturbative QCD and employ the leading twist contribution from the operator product expansion~\cite{Larin:1991tj,Baikov:2010je,Chetyrkin:2000yt}.
Further details can be found in Ref.~\cite{Feng:2020zdc}. A common representative value for the scale of $Q_{\mathrm{cut}}^2$ is 2 GeV$^2$. 
It is also feasible to vary this value to investigate potential systematic effects.\\

\noindent
{\bf Numerical analysis}: We use two lattice QCD gauge ensembles at the physical pion mass,
generated by RBC and UKQCD Collaborations using $2+1$-flavor domain wall fermion~\cite{RBC:2014ntl}. The ensemble
parameters are outlined in Table~\ref{tab:ensemble_parameter}. Both ensemble utilize Iwasaki + DSDR action. 
For each configuration we produce 1024 point-source and 1024 smeared-source propagators at random spatial-temporal locations and calculate the correlation function
$\langle \psi_{p}(t_f)J_\mu^{em}(x)J_\nu^{W,A}(y)\psi_{n}^\dagger(t_i)\rangle$ with
$t_f=\operatorname{max}\{t_x,t_y\}+\Delta t_f$ and
$t_i=\operatorname{min}\{t_x,t_y\}-\Delta t_i$ using the random sparsening-field technique~\cite{Li:2020hbj,Detmold:2019fbk}. Local vector and axial vector current operators 
are contracted with the renormalization factors quoted from Ref.~\cite{Feng:2021zek}.
We calculate all the connected insertions, discarding disconnected insertions which vanish under the flavor SU(3) limit.

\begin{table}[htbp]
\small
\centering
\begin{tabular}{cccclc}
\hline
\hline
Ensemble  & $m_\pi$ [MeV] & $L$ &  $T$ & $a^{-1}$ [GeV]&
$N_{\text{conf}}$ \\
\hline
24D  & $142.6(3)$ & $24$ & $64$ & $1.023(2)$ & 207   \\
32D-fine & $143.6(9)$ & $32$ & $64$ & $1.378(5)$ & 69 \\
\hline
\end{tabular}%
\caption{Ensembles used in this work. For each ensemble we list the pion mass $m_\pi$, 
the spatial and temporal extents, $L$ and $T$, 
the inverse of lattice
spacing $a^{-1}$~\cite{RBC:2023xqv}, the number
of configurations used, $N_{\text{conf}}$. The lattice spacing is determined using the mass of $\Omega$ baryon as input.}
\label{tab:ensemble_parameter}%
\end{table}%

\begin{figure}[htb]
\centering
\includegraphics[width=0.48\textwidth,angle=0]{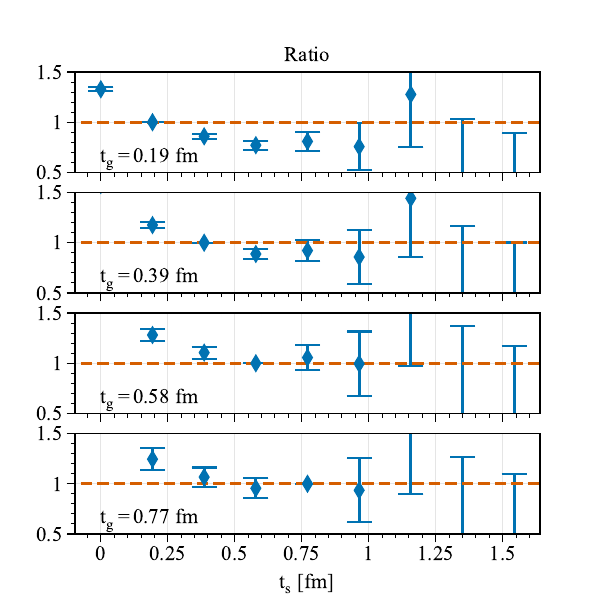}
\caption{The ratio defined in Eq.~(\ref{eq:ratio}) as a function of $t_s$. Here, we employ 24D as an illustrative example. $\Delta t_i+\Delta t_f$ is fixed at 0.77 fm.
}
\label{fig:ratio}
\end{figure}

To incorporate the LD contribution, the appropriate values for $t_g$ and $t_s$ must be determined. We calculate the LD contribution to $\Box_{\gamma
W}^{VA,\leq 2\mathrm{GeV}^2}$ using $M_n^{\text{LD}}(Q^2,t_s,t_g)$ as inputs, labelling the relevant part of the box contribution as $\Box_{\gamma
W}^{VA,\leq 2\mathrm{GeV}^2}(t_s,t_g)$. For small $t_g$ values, a visible contamination from excited states is anticipated.
To extend this analysis, we calculate $\Box_{\gamma
W}^{VA,\leq 2\mathrm{GeV}^2}(t_s,t_s)$ with $t_g=t_s$ for various $t_s$ values and construct a ratio
\begin{equation}
\label{eq:ratio}
\text{Ratio}=\frac{\Box_{\gamma
W}^{VA,\leq 2\mathrm{GeV}^2}(t_s,t_g)}{\Box_{\gamma
W}^{VA,\leq 2\mathrm{GeV}^2}(t_s,t_s)},
\end{equation}
as depicted in Fig.~\ref{fig:ratio}. Evidently, for $t_g$ below 0.39\,fm, the ratio is not fully consistent with 1 for $t_s\ge t_g$. 
However, at $t_g\ge 0.58$ fm, the LD contribution reconstructed using $H(t_s=t_g,\vec{x})$ and $H(t_s> t_g,\vec{x})$ agree well. 
This implies that at $t_g=0.58$\,fm, the ground state begins to dominate the hadronic function, and statistical deviations between the reconstruction at $t_s>t_g$ and $t_s=t_g$ are negligible.

\begin{figure}[htb]
\centering
\includegraphics[width=0.48\textwidth,angle=0]{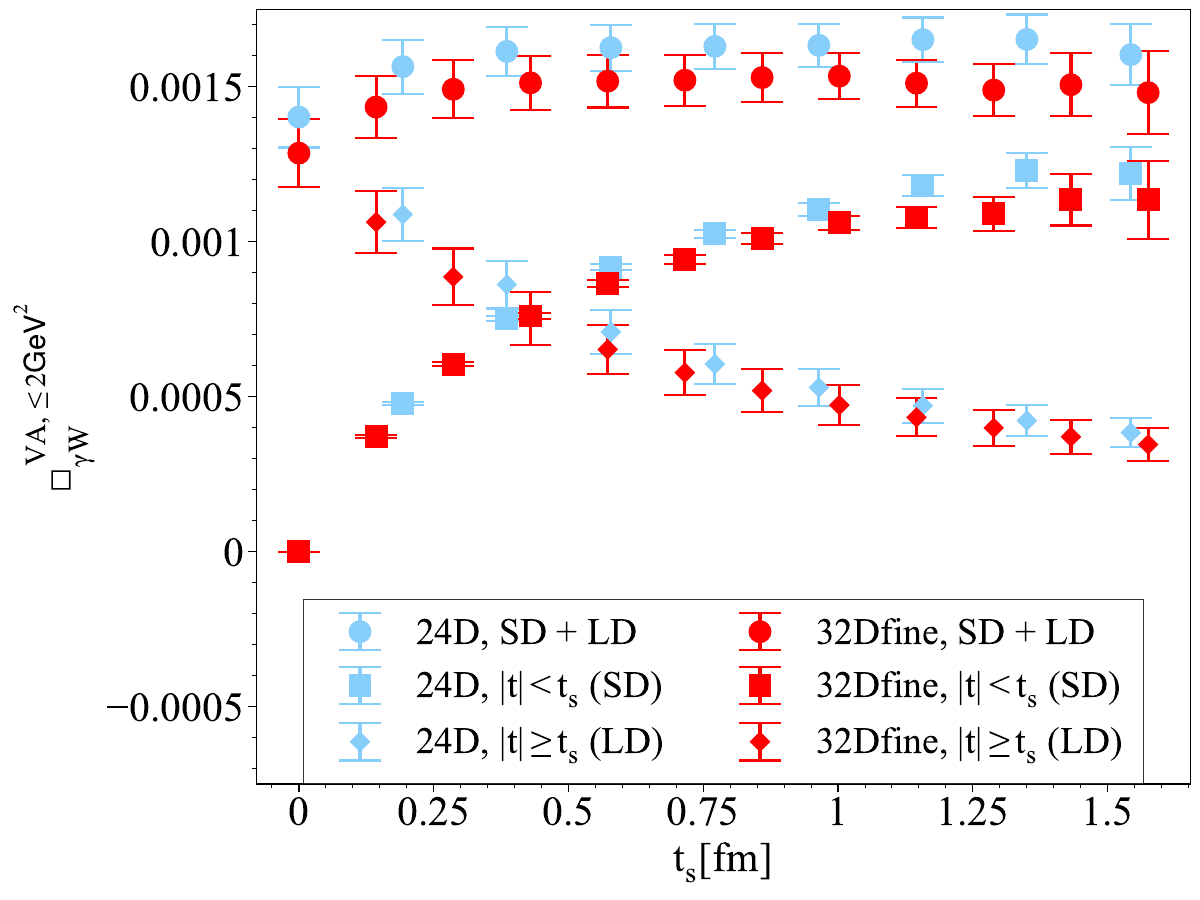}
\caption{SD, LD and total contributions to $\Box_{\gamma
W}^{VA,\leq 2\mathrm{GeV}^2}$ as a function of $t_s$. The IVR is performed at $t_g=0.58$ fm for 24D and $t_g=0.57$ fm for 32Dfine. $\Delta t_i+\Delta t_f$ is set at $\sim0.75$ fm for both ensembles.
}
\label{fig:SD_LD}
\end{figure}

To further verify the ground-state dominance, we present in Fig.~\ref{fig:SD_LD} the SD, LD and total contributions to $\Box_{\gamma
W}^{VA,\leq 2\mathrm{GeV}^2}$ for both the 24D and 32Dfine ensembles . The LD contribution is reconstructed utilizing the lattice data of $H(t_g,\vec{x})$ with $t_g\approx 0.6$ fm. 
When the SD and LD contributions are combined, a discernible plateau emerges for $t_s\gtrsim 0.6$ fm,
indicating that the influence of excited-state contributions beyond this specific time slice is notably mild.

\begin{figure}[htb]
\centering
\includegraphics[width=0.48\textwidth,angle=0]{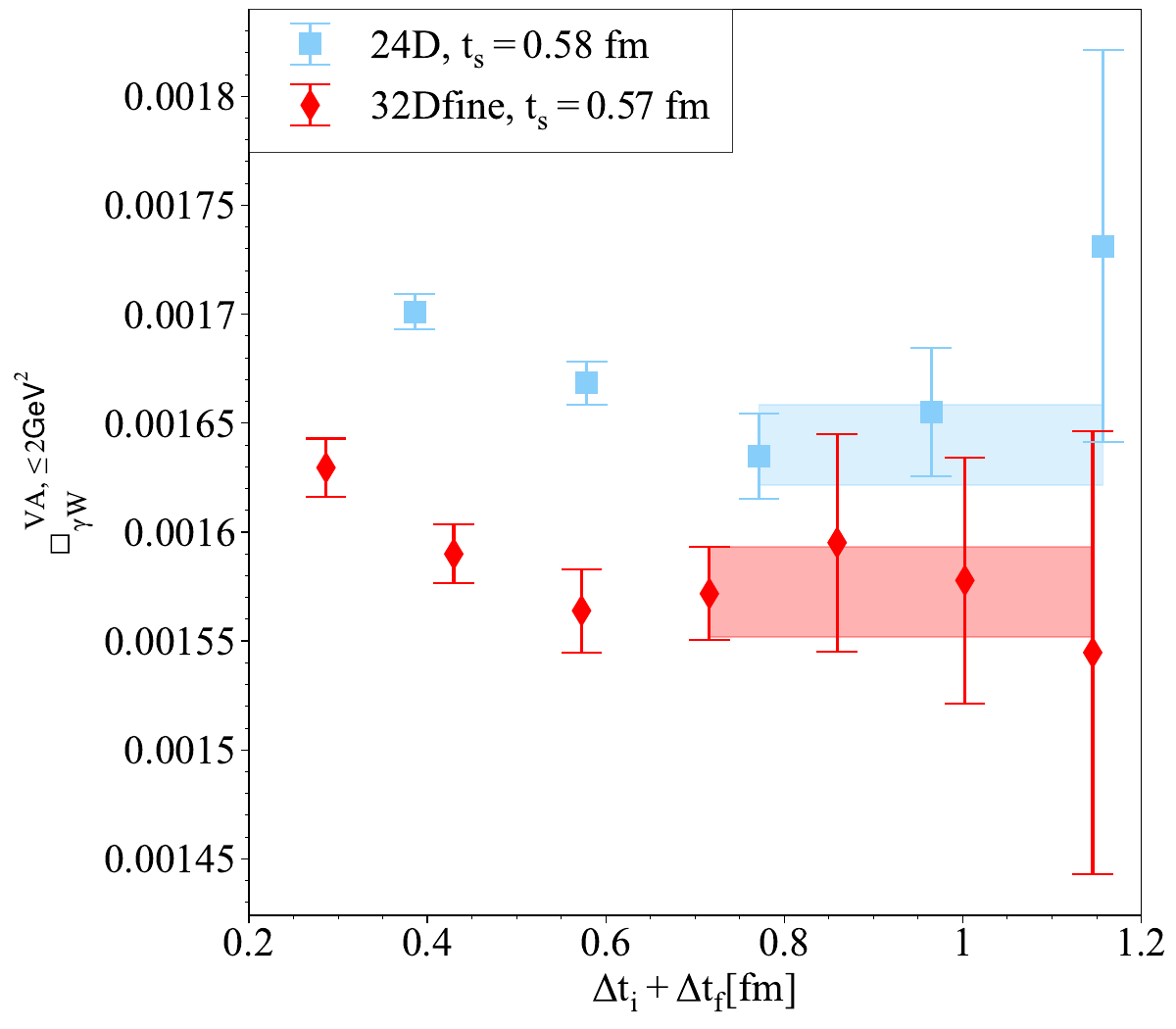}
\caption{$\Box_{\gamma
W}^{VA,\le 2 \mathrm{GeV}^2}$ for 24D and 32Dfine as a function of $\Delta t_i+\Delta t_f$. Here $t_s$ and $t_g$ are maintained as $t_s=t_g\approx 0.6$ fm.
}
\label{fig:dt_dep}
\end{figure}

\begin{figure}[htb]
\centering
\includegraphics[width=0.48\textwidth,angle=0]{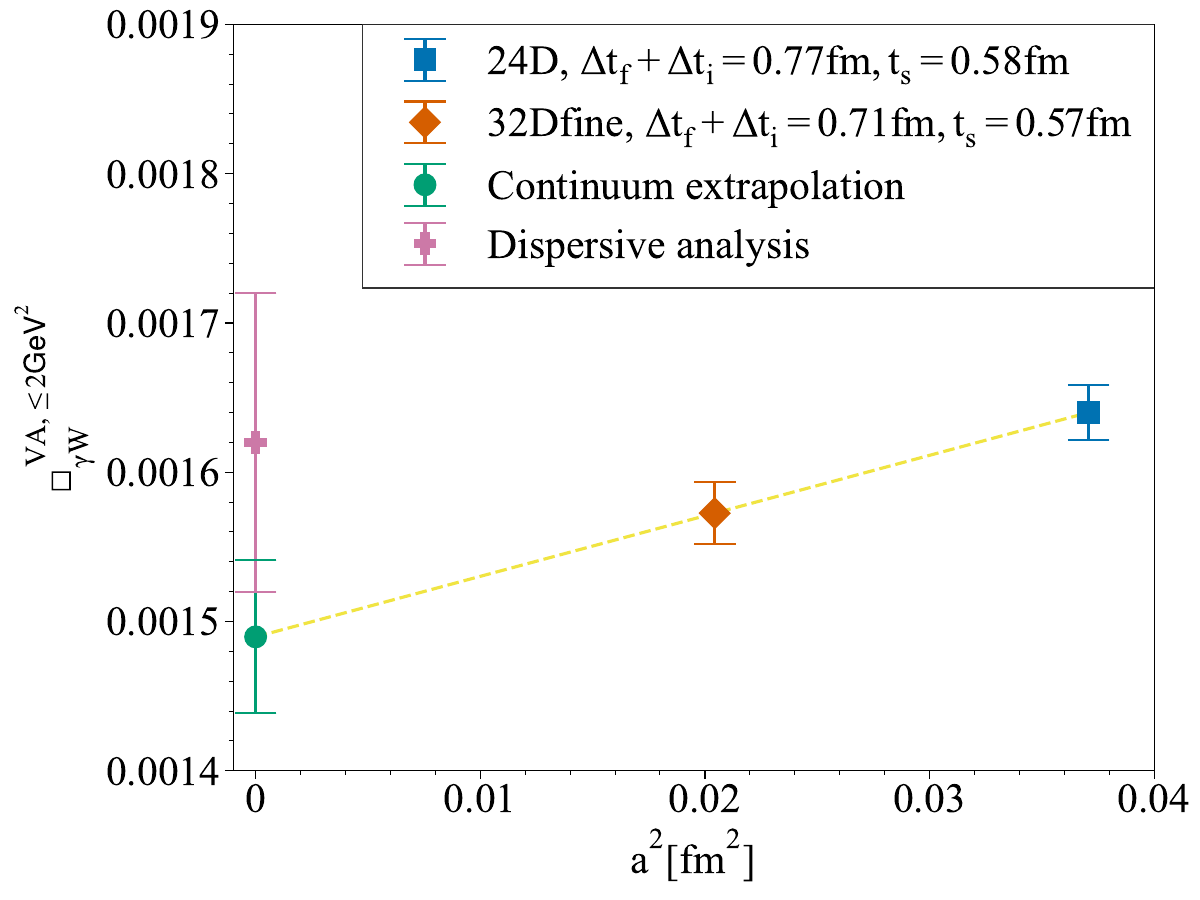}
\caption{Continuum extrapolation for $\Box_{\gamma
W}^{VA,\le 2 \mathrm{GeV}^2}$. The lattice result at the continuum limit is slightly lower than prediction from dispersive analysis.
}
\label{fig:continue}
\end{figure}

In Fig.~\ref{fig:dt_dep} we present a plot of $\Box_{\gamma
W}^{VA,\leq 2\mathrm{GeV}^2}$ as a function of $\Delta t_i+\Delta t_f$ utilizing $t_s=t_g\approx0.6$ fm for both 24D and 32Dfine. 
The maximal source-sink separation extends to to $\sim 1.8$ fm.
In our analysis, we note that when $\sqrt{Q^2}<0.2$ GeV, the substitution method's error is less than that of the direct method.
Thus, we adopt the substitution method for calculating the momentum integral for $\Box_{\gamma
W}^{VA,\le 2 \mathrm{GeV}^2}$ when $\sqrt{Q^2}<0.2$ GeV, and resort to the direct method for integrals
exceeding this threshold. For $\Delta t_i+\Delta t_f\gtrsim 0.7$ fm lattice results manifest a discernible plateau for both 24D and 32Dfine ensembles. By fitting to a constant, 
we obtain
$\Box_{\gamma
W}^{VA,\le 2 \mathrm{GeV}^2}$ values for each ensemble,  which in turn serve as the basis for a continuum extrapolation, as carried out in Fig.~\ref{fig:continue}, where we obtain
$\Box_{\gamma
W}^{VA,\leq 2\mathrm{GeV}^2}=1.490(51)\times 10^{-3}$
in the continuum limit. 
In the study of the pion decay~\cite{Feng:2020zdc}, we observed that the continuum extrapolation derived from the two Iwasaki ensembles with finer lattice spacings, 48I/64I, 
deviates from the outcomes of the 24D/32Dfine ensembles by $2.4\times 10^{-5}$. 
Considering that lattice discretization effects predominantly arise from short-distance effects in the loop momentum integral, we anticipate consistency between the pion and the nucleon.
In practice, we adopt a more conservative approach by assigning a twice-larger uncertainty, i.e. $4.8\times10^{-5}$, for the nucleon than for the pion. A future calculation
performed at finer lattice spacings would be beneficial for a more robust error estimate.
The other systematic effects including finite-volume effects (FV), excited-state contaminations (ES) and contributions from disconnected digrams (disc)~\cite{Gerardin:2023naa} are also estimated, with the details incorporated in the supplementary materials. Finally, we obtain
\begin{eqnarray}
\label{eq:box_lattice}
&&\Box_{\gamma
W}^{VA,\leq 2\mathrm{GeV}^2}
\nonumber\\
&=&1.490(51)_{\mathrm{stat}}(48)_{\mathrm{a}}(11)_{\mathrm{FV}}(12)_{\mathrm{ES}}(15)_{\mathrm{disc}}\times 10^{-3},
\nonumber\\
&=&1.490(73)\times10^{-3}.
\end{eqnarray}
This value is in agreement within 1$\sigma$ with the value $\Box_{\gamma
W}^{VA,\leq 2\mathrm{GeV}^2}=1.62(10)\times 10^{-3}$ obtained from dispersive analysis~\cite{Seng:2018yzq,Seng:2018qru}.

At the leading-twist level, the perturbative contribution to $\Box_{\gamma
W}^{VA,> 2 \mathrm{GeV}^2}$ aligns with that for the pion decay. 
Therefore, we utilize the 4-loop analysis detailed in Ref.~\cite{Feng:2020zdc} to deduce
\begin{equation}
\Box_{\gamma
W}^{VA,> 2\mathrm{GeV}^2}=2.159(6)_{\mathrm{HL}}(3)_{\mathrm{HT}}\times10^{-3},
\end{equation}
where the first error accounts for the higher-loop truncation effects. This is determined by contrasting the results from 4-loop and 3-loop perturbative calculations.  
The second error pertains to higher-twist truncation effect and is gauged through diagrams involving two currents positioned on distinct quark lines, thus containing solely higher-twist contributions.
Upon combining $\Box_{\gamma
W}^{VA,> 2 \mathrm{GeV}^2}$ and $\Box_{\gamma
W}^{VA,\le 2 \mathrm{GeV}^2}$, the final result becomes
\begin{equation}
\Box_{\gamma
W}^{VA}=3.65(7)_{\mathrm{lat}}(1)_{\mathrm{PT}}\times10^{-3}.
\end{equation}
To assess the $Q^2_{\text{cut}}$ dependence,
we also consider $Q^2_{\text{cut}}=1$ GeV$^2$ and $3$ GeV$^2$, with remarkably consistent results, as is seen from Table~\ref{tab:box_contribution}. 

\begin{table}[htbp]
\small
\centering
\begin{tabular}{c|c|c|c}
\hline
\hline
$Q^2_{\mathrm{cut}}$  & $\Box_{\gamma
W}^{VA,\le Q_{\mathrm{cut}}^2}$ & $\Box_{\gamma
W}^{VA,> Q_{\mathrm{cut}}^2}$  & $\Box_{\gamma
W}^{VA}$\\
\hline
1 GeV$^2$ & $ 1.32(7)\times10^{-3} $ & $2.31(2)\times10^{-3}$ & $ 3.63(7)\times 10^{-3} $ \\
2 GeV$^2$ & $1.49(7)\times 10^{-3}$ & $2.16(1)\times10^{-3}$ & $ 3.65(7)\times 10^{-3} $ \\
3 GeV$^2$ & $ 1.59(7)\times 10^{-3}$ & $2.06(1)\times10^{-3}$ & $ 3.65(7)\times 10^{-3} $ \\
\hline
\hline
\end{tabular}%
\caption{
Utilizing the scale $Q^2_{\mathrm{cut}}$ partition the integral range, we present
the contributions of $\Box_{\gamma W}^{VA,\le Q^2_{\mathrm{cut}}}$ from lattice QCD and
$\Box_{\gamma W}^{VA,>Q^2_{\mathrm{cut}}}$ from perturbation theory.}
\label{tab:box_contribution}%
\end{table}%

%\section{Results and conclusion}
\noindent
{\bf Results and conclusion}: According to the PDG~\cite{Workman:2022ynf}, precise values of $|V_{ud}|^2$ are obtained from superallowed nuclear and neutron beta decays~\cite{Towner:2010zz,Hardy:2020qwl,Czarnecki:2004cw,Marciano:2005ec} as
\begin{eqnarray}
&&\left|V_{u d}\right|^2=\frac{0.97154(22)_{\mathrm{exp}}(54)_{\mathrm{NS}}}{(1+\Delta_R^V)},\quad \text{superallowed},\\
&&\left|V_{u d}\right|^2=\frac{0.9728(6)_{\tau_n}(16)_{g_A}}{\left(1+\Delta_R^V\right)},\quad \text{free neutron}.
\end{eqnarray}
For superallowed decays, the first uncertainty stems from analyzing the half-lives of 15 precisely measured decays~\cite{Hardy:2020qwl}. Additionally, an uncertainty associated with the nuclear structure, $(54)_{\text{NS}}$, is quoted from Ref.~\cite{Gorchtein:2018fxl}. 
For neutron decays, the primary contributors to the uncertainty of $|V_{ud}|$ are of the experimental origin, namely the neutron lifetime $\tau_n=878.4(5)$~s
and the ratio of axial-vector to vector couplings, $g_A=1.2754(13)$~\cite{Workman:2022ynf}. It is worth noting that due to the SM radiative corrections $g_A$ deviates from the axial charge $\mathring{g}_A$ calculated in lattice QCD.
$\Delta_R^V$ denotes a universal, nuclear-structure-independent electroweak radiative correction. 
In the framework proposed by Sirlin~\cite{Sirlin:1977sv}, $\Delta_R^V$ is given by 
\begin{equation}
\Delta_R^V=\frac{\alpha}{2 \pi}\left[3 \ln \frac{M_Z}{m_p}+\ln \frac{M_Z}{M_W}+\tilde{a}_g\right]+\delta_{\mathrm{HO}}^{\mathrm{QED}}+2 \Box_{\gamma W}^{V A}
\end{equation}
with $\tilde{a}_g=-0.083$ representing the $\mathcal{O}(\alpha_s)$ QCD correction to all one-loop diagrams except the axial $\gamma W$ box. Additionally, $\delta_{\mathrm{HO}}^{\mathrm{QED}} = 0.00109(10)$~\cite{Czarnecki:2004cw} summarizes the leading-log higher-order QED effects, which can be accounted for via the running $\alpha_e$ (see an updated discussion in Ref.~\cite{Cirigliano:2023fnz}). 
Using the lattice results for $\Box_{\gamma W}^{V A}$, we obtain
\begin{eqnarray}
&&\Delta_R^V=0.02439(15)_{\mathrm{lat}}(10)_{\mathrm{HO}},\\
&&|V_{ud}|=0.97386(11)_{\mathrm{exp.}}(9)_{\mathrm{RC}}(27)_{\mathrm{NS}}
\label{eq:updated_Vud}
\end{eqnarray}
for superallowed beta decays. 
Upon entering $|V_{ud}|$, the uncertainties 
from both lattice results and $\delta_{\mathrm{HO}}^{\mathrm{QED}}$ contribute collectively as 
$(9)_{\mathrm{RC}}$, with the lattice uncertainty accounting for 83\%.
Likewise, for free neutron decay,
\begin{equation}
|V_{ud}|=0.9745(3)_{\tau_n}(8)_{g_A}(1)_{\mathrm{RC}}.
\end{equation}
We note that the primary source of uncertainty is attributed to the experimental measurements of $g_A$.
In this study, we also calculate the $\gamma W$ box correction to $g_A$ using the formula provided in~\cite{Hayen:2020cxh}
\begin{equation}
\label{eq:corr_to_gA}
g_A=\mathring{g}_A\left[1+\Box_{\gamma W}^{VV}-\Box_{\gamma W}^{VA}\right] + \text{f.f.}\,.
\end{equation}
Interestingly, the vector $\gamma W$-box term largely cancels the axial one, resulting in a value of $\Box_{\gamma W}^{VV}-\Box_{\gamma W}^{VA}=0.07(11)\times10^{-3}$ that aligns with zero. This outcome also agree with the value of $0.13(13)\times10^{-3}$ derived from dispersive analysis~\cite{Gorchtein:2021fce}. It is essential to emphasize that the $\gamma W$ box correction to $g_A$ doesn't constitute the dominant contribution. Ref.~\cite{Cirigliano:2022hob} suggests that the most substantial radiative correction originates from vertex corrections, and may approach a percent level. To address this possibility on the lattice, one would need to include, e.g., the radiative corrections to the axial form factors (``f.f.'' in the equation above). This, in turn, requires computing 5-point correlation functions on the lattice.

To conclude, in this work we perform the first lattice QCD calculation of the $\gamma W$-box contribution 
to the superallowed nuclear and free neutron beta decays. 
As depicted in Fig.~\ref{fig:continue}, lattice results for both ensembles are consistent with the outcome obtained from dispersive analysis. 
The only distinction arises in the slightly smaller value exhibited by the continuum-extrapolated result in comparison to the dispersive analysis.
It is always advisable to have future independent check using different discretizations of QCD.
Using the updated $|V_{ud}|$ as provided in Eq.~(\ref{eq:updated_Vud}) and combining it with $|V_{us}|=0.2243(8)$ from PDG \cite{Workman:2022ynf}, we obtain
\begin{equation}
|V_{ud}|^2 + |V_{us}|^2 + |V_{ub}|^2 = 0.9987(6)_{V_{ud}}(4)_{V_{us}}.
\end{equation}
This result exhibits $1.8\sigma$ tension with CKM unitarity.

\begin{acknowledgments}
{\bf Acknowledgments} -- X.F. and L.C.J. gratefully acknowledge many helpful discussions with our colleagues from the
RBC-UKQCD Collaborations.
X.F., P.X.M. and Z.L.Z. were supported in part by NSFC of China under Grant No. 12125501, No. 12070131001, and No. 12141501, 
and National Key Research and Development Program of China under No. 2020YFA0406400.
P.X.M. and Z.L.Z. were supported in part by NSFC of China under Grant No. 12293060 and No. 12293063.
L.C.J. acknowledges support by DOE Office of Science Early Career Award No. DE-SC0021147 and DOE Award No. DE-SC0010339. The work of M.G. is supported in part by EU Horizon 2020 research and innovation programme, STRONG-2020 project
under grant agreement No 824093, and by the Deutsche Forschungsgemeinschaft (DFG) under the grant agreement GO 2604/3-1. 
K.F.L. and B.G.W. are partially supported by the DOE Grant No. DE-SC0013065 and No. DE-AC05-06OR23177.
The work of C.-Y.S. is supported in
part by the U.S. Department of Energy (DOE), Office of Science, Office of Nuclear Physics, under the FRIB Theory Alliance award DE-SC0013617, by the DOE grant DE-FG02-97ER41014, and by the DOE Topical Collaboration ``Nuclear Theory for New Physics'', award No.
DE-SC0023663. The work of B.G.W. is supported in part by the DOE Office of Science, Office of Nuclear Physics under the umbrella of the Quark-Gluon Tomography (QGT) Topical Collaboration with Award DE-SC0023646.
The research reported in this work was carried out using the computing facilities at Chinese National Supercomputer Center in Tianjin. 
It also made use of computing and long-term storage facilities of the USQCD Collaboration, which are funded by the Office of Science of the U.S. Department of Energy.
\end{acknowledgments}

\bibliography{ref}

\clearpage

\setcounter{page}{1}
\renewcommand{\thepage}{Supplementary Information -- S\arabic{page}}
\setcounter{table}{0}
\renewcommand{\thetable}{S\,\Roman{table}}
\setcounter{equation}{0}
\renewcommand{\theequation}{S\,\arabic{equation}}
\setcounter{figure}{0}
\renewcommand{\thefigure}{S\,\arabic{figure}}

\section{Supplementary Material}

\subsection{Quark contractions for nucleon 4-point correlation functions}

\begin{figure}[htb]
\centering
\includegraphics[width=0.48\textwidth,angle=0]{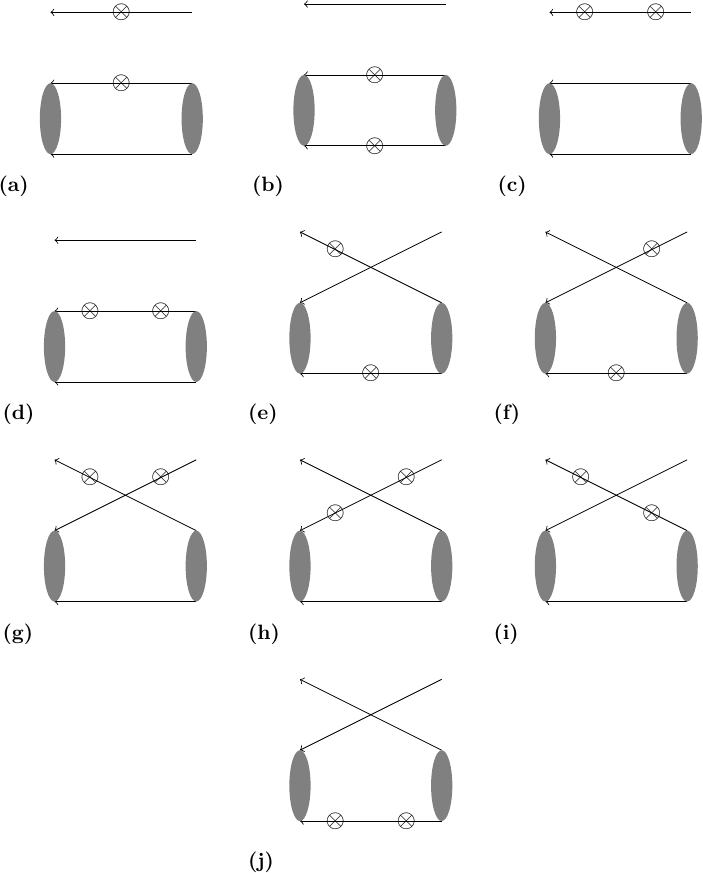}
\caption{8 out of the 10 quark contraction types are pertinent to the neutron $\gamma W$-box diagrams. 
The symbols of $\otimes$ indicate the insertions of the vector or axial-vector current.}
\label{fig:contraction}
\end{figure}

The connected insertions encompass 10 distinct contraction types, as illustrated in Fig.~\ref{fig:contraction}. 
Notably, types (b) and (d) do not contribute to the axial-vector and vector $\gamma W$-box diagrams. 
This is because the quark current associated with the $W$ boson, altering the isospin, cannot be inserted between isospin-0 diquark blocks.

\subsection{Demonstration of Eq.~(\ref{eq:MnLD_direct})}

We follow the derivation presented in Ref.~\cite{Feng:2020zdc} to express $M_n(Q^2)$ as
\begin{equation}
\label{eq:weighted_M}
M_n(Q^2)=-\frac{i}{3m_H}\int_{-\sqrt{Q^2}}^{\sqrt{Q^2}}\frac{dQ_0}{\pi}\frac{|\vec{Q}|}{Q^2}\epsilon_{\mu\nu\alpha 0}Q_{\alpha}T_{\mu\nu}^{VA}(Q).
\end{equation}
Here, the hadronic tensor $T_{\mu\nu}^{VA}(Q)$ is defined as
\begin{eqnarray}
T_{\mu\nu}^{VA}(Q)&=&\frac{1}{2}\int dt\,e^{-iQ_0t}\,\tilde{\mathcal{H}}_{\mu\nu}^{VA}(t,\vec{Q})
\nonumber\\
&=&\frac{1}{2}\int dt\,e^{-iQ_0t}\int d^3\vec{x}\,e^{-i\vec{Q}\cdot\vec{x}}\mathcal{H}_{\mu\nu}^{VA}(t,\vec{x}).
\end{eqnarray}
The tensor $T_{\mu\nu}^{VA}(Q)$ is split into long-distance and short-distance parts
\begin{eqnarray}
T_{\mu\nu}^{VA}(Q)&=&T_{\mu\nu}^{\mathrm{SD}}(t_s,Q)+T_{\mu\nu}^{\mathrm{LD}}(t_s,Q)
\nonumber\\
&=&\frac{1}{2}\int_{-t_s}^{t_s} dt\,e^{-iQ_0t}
\tilde{\mathcal{H}}_{\mu\nu}^{VA}(t,\vec{Q})
\nonumber\\
&+&\frac{1}{2}
\left(\int_{-\infty}^{-t_s}dt+\int_{t_s}^\infty dt\right)e^{-iQ_0t}\tilde{\mathcal{H}}_{\mu\nu}^{VA}(t,\vec{Q}).
\nonumber\\
\end{eqnarray}
At $|t|\ge t_g$, $\tilde{\mathcal{H}}_{\mu\nu}^{VA}(t,\vec{Q})$ can be expressed in terms of $\tilde{\mathcal{H}}_{\mu\nu}^{VA}(\pm t_g,\vec{Q})$ as
\begin{equation}
\label{eq:gs-dominance}
\tilde{\mathcal{H}}_{\mu\nu}^{VA}(t,\vec{Q})=
\begin{cases}
\tilde{\mathcal{H}}_{\mu\nu}^{VA}(t_g,\vec{Q})\,e^{-(E_{\vec{Q}}-m_N)(t-t_g)}, & \mbox{for }t\ge t_g\\
\tilde{\mathcal{H}}_{\mu\nu}^{VA}(-t_g,\vec{Q})\,e^{(E_{\vec{Q}}-m_N)(t+t_g)}, & \mbox{for }t\le -t_g
\end{cases}.
\end{equation}
Utilizing Eq.~(\ref{eq:gs-dominance}), $T_{\mu\nu}^{\mathrm{LD}}(t_s,t_g,Q)$ is expressed as
\begin{eqnarray}
T_{\mu\nu}^{\mathrm{LD}}(t_s,t_g,Q)&=&\frac{1}{2}\frac{e^{-iQ_0t_s}e^{-(E_{\vec{Q}}-m_N)(t_s-t_g)}}{E_{\vec{Q}}-m_N+iQ_0}\tilde{\mathcal{H}}_{\mu\nu}^{VA}(t_g,\vec{Q})
\nonumber\\
&+&\frac{1}{2}\frac{e^{iQ_0t_s}e^{(E_{\vec{Q}}-m_N)(-t_s+t_g)}}{E_{\vec{Q}}-m_N-iQ_0}\tilde{\mathcal{H}}_{\mu\nu}^{VA}(-t_g,\vec{Q}).
\nonumber\\
\end{eqnarray}
Taking the real part, we obtain
\begin{eqnarray}
T_{\mu\nu}^{\mathrm{LD}}(t_s,t_g,Q)&=&\operatorname{Re}
\left(\frac{e^{-iQ_0t_s}}{E_{\vec{Q}}-m_N+iQ_0}\right)e^{-(E_{\vec{Q}}-m_N)(t_s-t_g)}
\nonumber\\
&\times&\frac{1}{2}
\left(\tilde{\mathcal{H}}_{\mu\nu}^{VA}(t_g,\vec{Q})+\tilde{\mathcal{H}}_{\mu\nu}^{VA}(-t_g,\vec{Q})\right).
\end{eqnarray}
Inserting $T_{\mu\nu}^{\mathrm{LD}}(t_s,t_g,Q)$ into Eq.~(\ref{eq:weighted_M}), we derive
\begin{eqnarray}
\label{eq:M_LD_rewrite}
M_n^{\mathrm{LD}}(Q^2,t_s,t_g)&=&-\frac{i}{6m_H}\int_{-\sqrt{Q^2}}^{\sqrt{Q^2}}\frac{dQ_0}{\pi}\frac{|\vec{Q}|}{Q^2}
\nonumber\\
&&\times  \operatorname{Re}
\left(\frac{e^{-iQ_0t_s}}{E_{\vec{Q}}-m_N+iQ_0}\right)e^{-(E_{\vec{Q}}-m_N)(t_s-t_g)}
\nonumber\\
&&\times
\epsilon_{\mu\nu\alpha 0}Q_{\alpha}\left(\tilde{\mathcal{H}}_{\mu\nu}^{VA}(t_g,\vec{Q})+\tilde{\mathcal{H}}_{\mu\nu}^{VA}(-t_g,\vec{Q})\right).
\nonumber\\
\end{eqnarray}
The third line of Eq.~(\ref{eq:M_LD_rewrite}) is expressed in terms of $\bar{H}(t_g,\vec{x})$ through
\begin{eqnarray}
\label{eq:third_line}
&&\epsilon_{\mu\nu\alpha 0}Q_{\alpha}\left(\tilde{\mathcal{H}}_{\mu\nu}^{VA}(t_g,\vec{Q})+\tilde{\mathcal{H}}_{\mu\nu}^{VA}(-t_g,\vec{Q})\right)
\nonumber\\
&=&-i\epsilon_{\mu\nu\alpha 0}\int d^3\vec{x}\,e^{-i\vec{Q}\cdot\vec{x}}\partial_\alpha\left(\mathcal{H}_{\mu\nu}^{VA}(t_g,\vec{x})+\mathcal{H}_{\mu\nu}^{VA}(-t_g,\vec{x})\right)
\nonumber\\
&=&-i\epsilon_{\mu\nu\alpha 0}\int d^3\vec{x}\,j_0\left(|\vec{Q}||\vec{x}|\right)\partial_\alpha\left(\mathcal{H}_{\mu\nu}^{VA}(t_g,\vec{x})+\mathcal{H}_{\mu\nu}^{VA}(-t_g,\vec{x})\right)
\nonumber\\
&=&-i\int d^3\vec{x}\,\frac{|\vec{Q}|}{|\vec{x}|}j_1\left(|\vec{Q}||\vec{x}|\right) \times 2\bar{H}(t_g,\vec{x}).
\end{eqnarray}
Inserting Eq.~(\ref{eq:third_line}) into Eq.~(\ref{eq:M_LD_rewrite}), we arrive at Eq.~(\ref{eq:MnLD_direct}).

\subsection{Demonstration of Eq.~(\ref{eq:low_Q_relation})}

Here we demonstrate that once the ground-state dominance is satisfied at $|t|\ge t_g$, the spatial summation of the hadronic function $H(t,\vec{x})$ can be written in terms of ${\mathring g}_A$, ${\mathring \mu}_p$ and ${\mathring \mu}_n$. We start with the expression
\begin{eqnarray}
&&\epsilon_{\mu\nu\alpha0}Q_\alpha \tilde{\mathcal{H}}_{\mu\nu}^{VA}(t\ge t_g,\vec{Q})
\nonumber\\
&=&\epsilon_{\mu\nu\alpha0}Q_\alpha \int d^3\vec{x}\,e^{-i\vec{Q}\cdot\vec{x}}\mathcal{H}_{\mu\nu}^{VA}(t\ge t_g,\vec{x})
\nonumber\\
&=&\epsilon_{\mu\nu\alpha0}Q_\alpha\frac{e^{(m_N-E_{\vec{Q}})t}}{2E_{\vec{Q}}}\times
\nonumber\\
&&\frac{1}{2}\operatorname{Tr}\left[m_N(1+\gamma_0)\mathcal{V}_{\mu}
\left(E_{\vec{Q}}\gamma_0-i\vec{Q}\cdot\vec{\gamma}+m_N\right)\mathcal{A}_{\nu}\right],
\nonumber\\
\end{eqnarray}
where 
\begin{eqnarray}
\mathcal{V}_{\mu}&=&\gamma_\mu F_1(Q^2)-\frac{\sigma_{\mu\lambda}Q_\lambda}{2m_N}F_2(Q^2),
\nonumber\\
\mathcal{A}_\nu&=&\gamma_\nu\gamma_5 G_A(Q^2)+\frac{Q_\nu}{2m_N} \gamma_5 \tilde{G}_P(Q^2).
\end{eqnarray}
In the small $|\vec{Q}|$ limit, the above expression can be simplified as
\begin{eqnarray}
\label{eq:small_Q}
&&\lim_{|\vec{Q}|\to0}\epsilon_{\mu\nu\alpha0}Q_\alpha \tilde{\mathcal{H}}_{\mu\nu}^{VA}(t\ge t_g,\vec{Q})
\nonumber\\
&=&-\frac{i}{4}\epsilon_{\mu\nu\alpha0}Q_\alpha Q_\lambda
G_M(Q^2)G_A(Q^2)\operatorname{Tr}[\gamma_0\gamma_\mu\gamma_\lambda\gamma_\nu\gamma_5]
\nonumber\\
&&+\mathcal{O}(|\vec{Q}|^3)
\nonumber\\
&=&2i|\vec{Q}|^2G_M(0)G_A(0)+\mathcal{O}(|\vec{Q}|^3),
\end{eqnarray}
where $G_A(0)={\mathring g}_A$. $G_M(Q^2)=F_1(Q^2)+F_2(Q^2)$ is the magnetic form factor. For $t\ge t_g$, we have $G_M(0)={\mathring \mu}_p$. For $t\le -t_g$, the expression is similar as
Eq.~(\ref{eq:small_Q}), albeit with $G_M(0)={\mathring \mu}_n$. We thus have
\begin{eqnarray}
&&\lim_{|\vec{Q}|\to0}\epsilon_{\mu\nu\alpha0}Q_\alpha \left[\tilde{\mathcal{H}}_{\mu\nu}^{VA}(t_g,\vec{Q})+\tilde{\mathcal{H}}_{\mu\nu}^{VA}(-t_g,\vec{Q})\right]
\nonumber\\
&=&2i|\vec{Q}|^2{\mathring g}_A({\mathring \mu}_p+{\mathring \mu}_n)+\mathcal{O}(|\vec{Q}|^3)
\end{eqnarray}

On the other hand, we have
\begin{eqnarray}
&&\epsilon_{\mu\nu\alpha0}Q_\alpha \tilde{\mathcal{H}}_{\mu\nu}^{VA}(t\ge t_g,\vec{Q})
\nonumber\\
&=&\epsilon_{\mu\nu\alpha0}\int d^3\vec{x}\,e^{-i\vec{Q}\cdot\vec{x}}(-i\partial_\alpha)\mathcal{H}_{\mu\nu}^{VA}(t\ge t_g,\vec{x})
\nonumber\\
&=&\epsilon_{\mu\nu\alpha0}\int d^3\vec{x}\,j_0(|\vec{Q}||\vec{x}|)(-i\partial_\alpha)\mathcal{H}_{\mu\nu}^{VA}(t\ge t_g,\vec{x})
\nonumber\\
&=&-i\int d^3\vec{x}\,\frac{j_1(|Q||\vec{x}|)}{|\vec{x}|}|\vec{Q}|H(t\ge t_g,\vec{x})
\end{eqnarray}
In the small $\vec{Q}$ limit, it yields
\begin{eqnarray}
\label{eq:small_Q1}
&&\lim_{|\vec{Q}|\to0}\epsilon_{\mu\nu\alpha0}Q_\alpha \left[\tilde{\mathcal{H}}_{\mu\nu}^{VA}(t_g,\vec{Q})+\tilde{\mathcal{H}}_{\mu\nu}^{VA}(-t_g,\vec{Q})\right]
\nonumber\\
&=&-2i\int d^3\vec{x}\,\frac{|\vec{Q}|^2}{3}\bar{H}(t_g,\vec{x})+\mathcal{O}(|\vec{Q}|^3).
\end{eqnarray}
Combining Eqs.~(\ref{eq:small_Q}) and (\ref{eq:small_Q1}), we obtain
\begin{equation}
\int d^3\vec{x}\,\bar{H}(t_g,\vec{x})=-3{\mathring g}_A({\mathring \mu}_p+{\mathring \mu}_n).
\end{equation}
In our calculation, we find that the lattice results for the 24D, 32Dfine ensembles, and the continuum extrapolation are all consistent with the PDG value, as depicted in Table~\ref{tab:zero_mom}.

\begin{table}[htbp]
 \centering
 \begin{tabular}{ccccc}
  \hline
  \hline
   & 24D & 32Dfine & Cont. & PDG \\
  \hline
  $-3 g_A (\mu_p+\mu_n)$ & $-3.31(49)$ & $-3.02(53)$ & $-2.65(1.31)$ & $-3.366(3)$\\
\hline
 \end{tabular}%
 \caption{The lattice results of $\int d^3\vec{x}\,\bar{H}(t_g,\vec{x})$ for the 24D, 32Dfine ensembles, and the continuum extrapolation contrasted with the PDG value of $-3g_A(\mu_p+\mu_n)$. Here, $t_g$ takes a value of $\sim0.6$ fm.}
 \label{tab:zero_mom}
\end{table}%

\subsection{Demonstration of the necessity of using the IVR method}

\begin{figure}[htb]
\centering
\includegraphics[width=0.48\textwidth,angle=0]{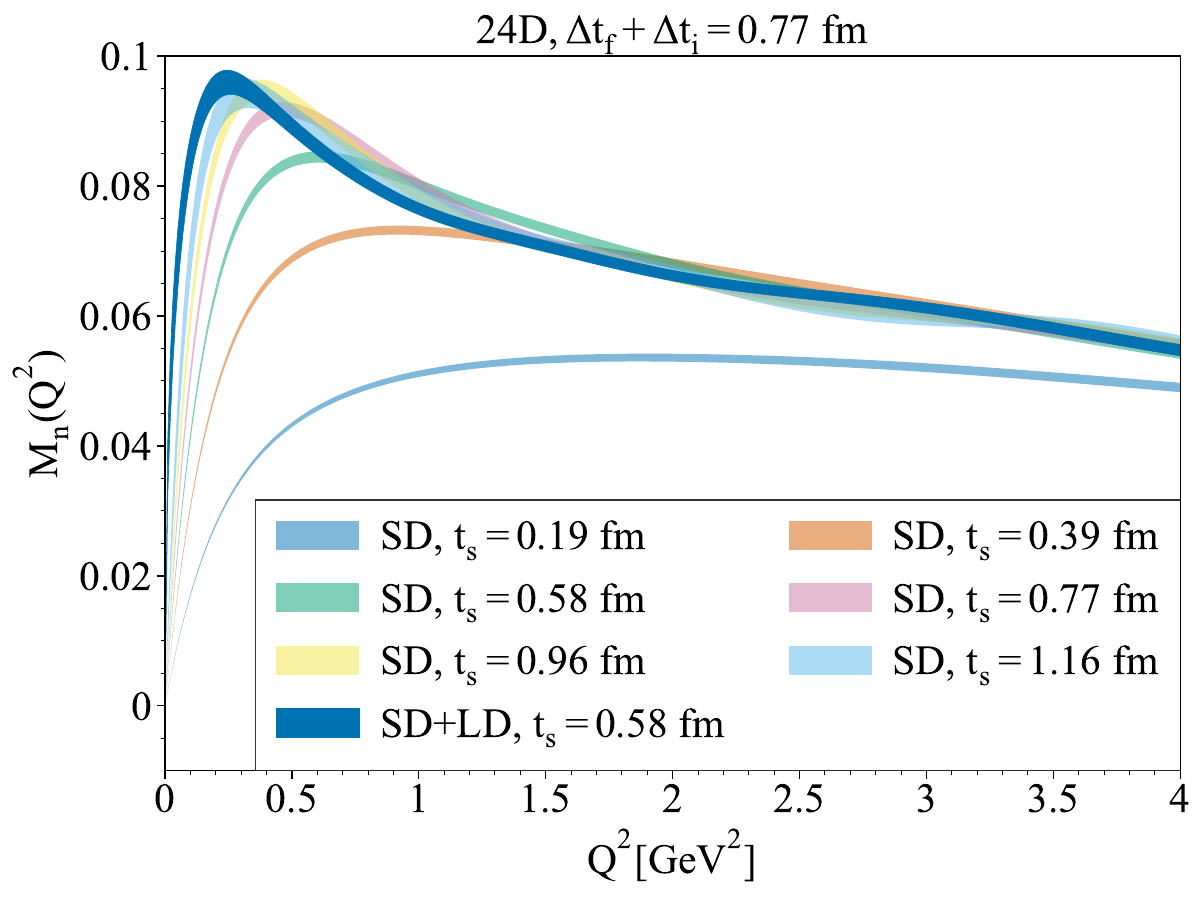}
\caption{SD and LD contributions to $M_n(Q^2)$ as a function of $Q^2$ with various choices of $t_s$ for 24D. $\Delta t_i+\Delta t_f$ is set at 0.77 fm. The error band denoted as ``SD+LD'' is the lattice result which incorporates the reconstruction of the LD contributions using IVR and $t_g=t_s=0.58$ fm.
}
\label{fig:msd_qsq}
\end{figure}

To demonstrate the necessity of using the IVR method in our calculation, we use the ensemble 24D as an example and present in Fig.~\ref{fig:msd_qsq} the results of $M_n^{\text{SD}}(Q^2, t_s)$ as a function of $Q^2$ 
for different values of $t_s$. 
Notably, even when increasing $t_s$ to 1.16\,fm while maintaining $\Delta t_i+\Delta t_f$ fixed at 0.77\,fm (resulting in a total source-sink separation of 1.93 fm), sizeable temporal truncation effects persist.

\subsection{Comparison between the two approaches for determining $M_n(Q^2)$}

\begin{figure}[htb]
\centering
\includegraphics[width=0.48\textwidth,angle=0]{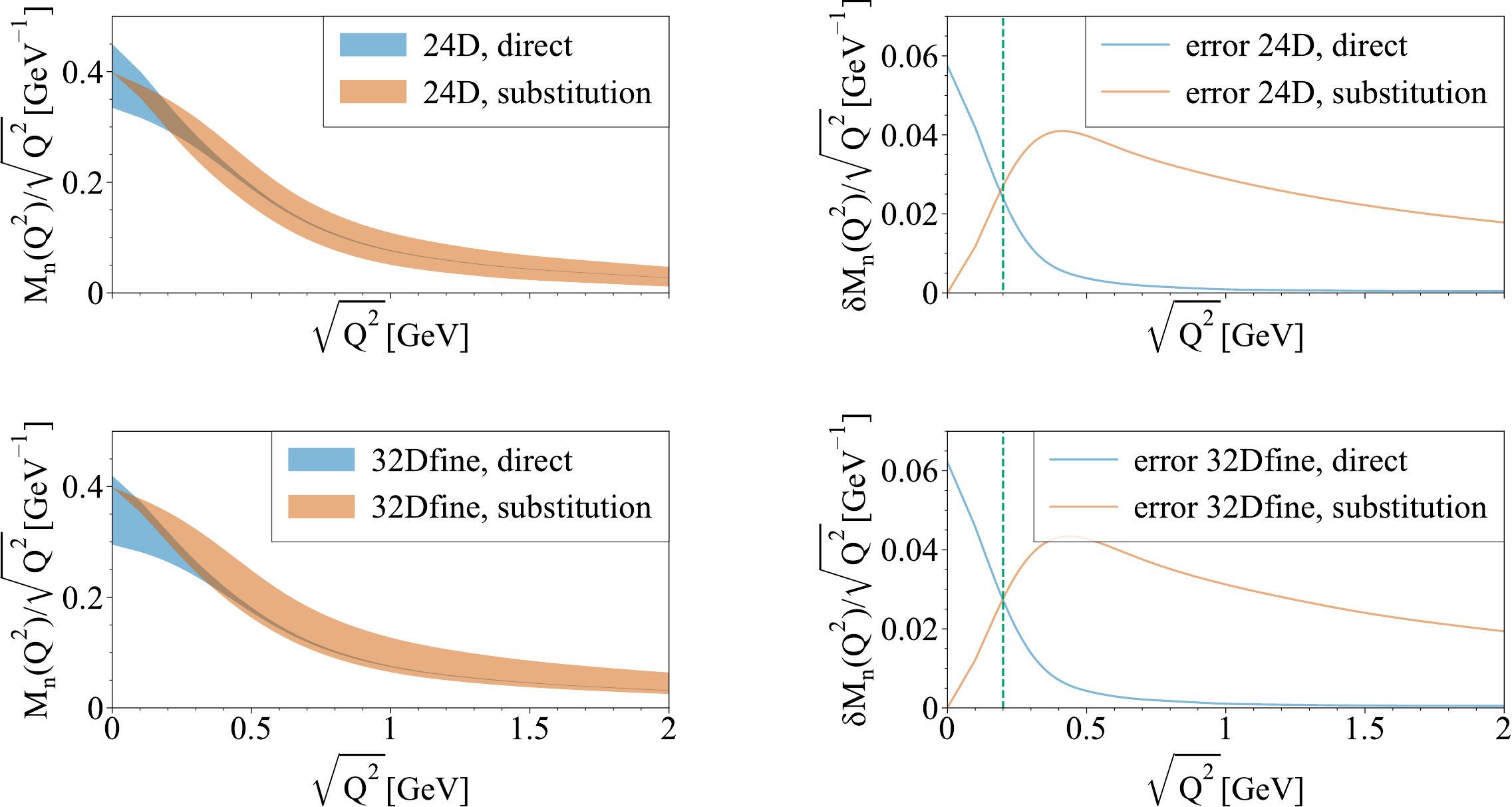}
\caption{In the left panel, we present $M_n(Q^2)/\sqrt{Q^2}$ as a function of $\sqrt{Q^2}$ for both the 24D and 32Dfine ensembles. Here $\Delta t_i+\Delta t_f$ is maintained at $\sim0.75$ fm, while $t_s=t_g$ is set at $\sim0.6$ fm for both ensembles. Meanwhile, the right panel depicts the uncertainty $\delta M_n(Q^2)/\sqrt{Q^2}$ as a function of $\sqrt{Q^2}$. There exists a uniform scale $(\sqrt{Q^2})_0\approx0.2$ GeV, marked by the dashed line, that when $\sqrt{Q^2}<(\sqrt{Q^2})_0$ the substitution method exhibits superiority, while beyond $(\sqrt{Q^2})_0$, the direct method holds the advantage.
}
\label{fig:comparison_direct_substitution}
\end{figure}

In our analysis, we employ two distinct approaches: one utilizes Eq.~(\ref{eq:MnLD_direct}), while the other employs Eq.~(\ref{eq:MnLD_substitution}) to determine $M_n^{\text{LD}}$. 
As depicted in Fig.~\ref{fig:comparison_direct_substitution}, both approaches yield results that are in alignment for the total contribution $M_n(Q^2)$.
For both 24D and 32Dfine, there exists a uniform scale $(\sqrt{Q^2})_0\approx0.2$ GeV. When $\sqrt{Q^2}<(\sqrt{Q^2})_0$, the error resulting from the substitution method is smaller than that of the direct method. Conversely, 
for $\sqrt{Q^2}>(\sqrt{Q^2})_0$, the reverse is true. Consequently, we adopt the substitution method to compute the integral when $\sqrt{Q^2}<(\sqrt{Q^2})_0$ and the direct method for integrals beyond that threshold.

\subsection{Comparison with dispersive analysis and perturbation theory}

\begin{figure}[htb]
     \centering
         \includegraphics[width=0.48\textwidth,angle=0]{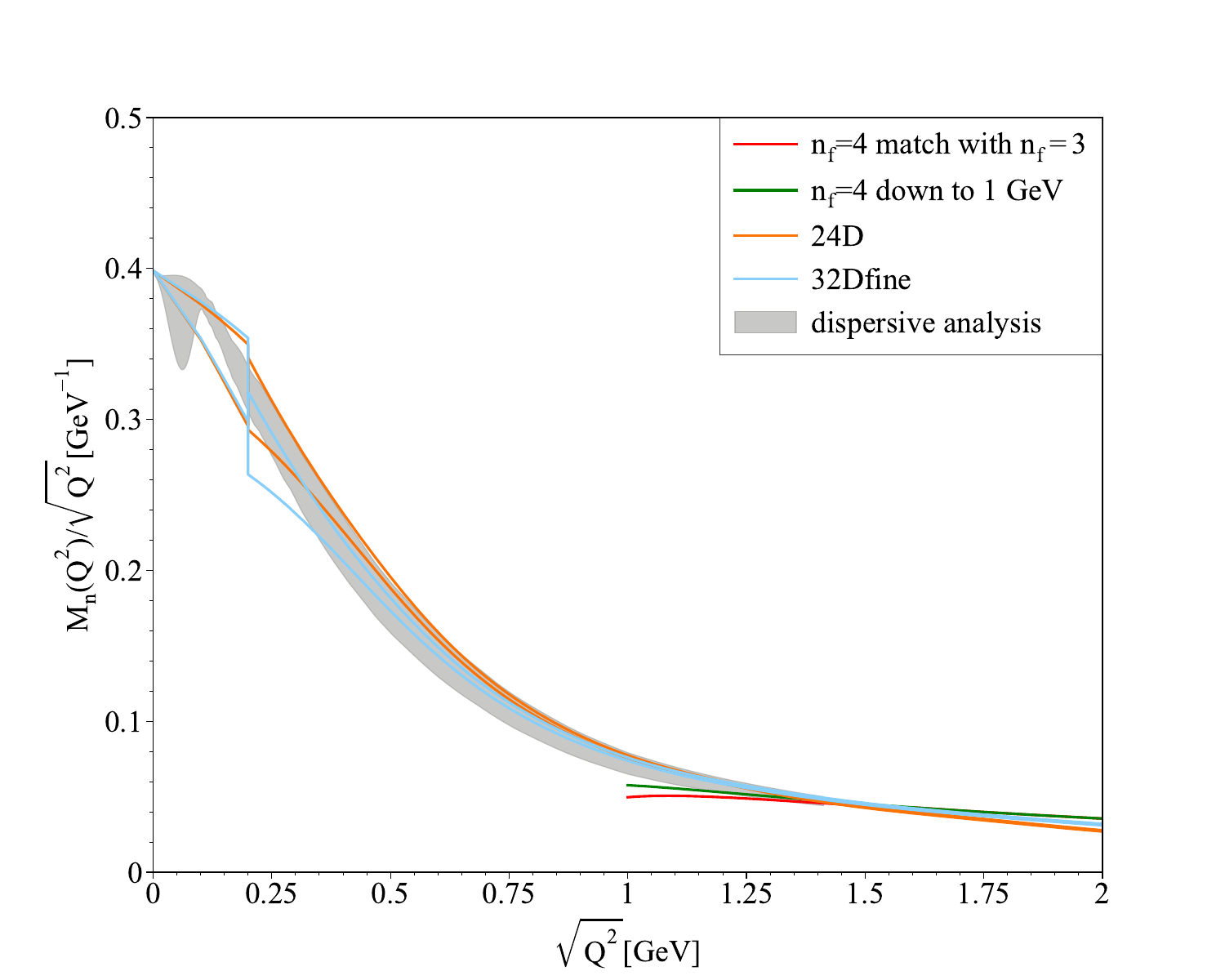}
         \caption{$M_n(Q^2)/\sqrt{Q^2}$ as a function of $\sqrt{Q^2}$. The parameters for $\Delta t_i+\Delta t_f$, $t_s$ and $t_g$ are maintained at the same values as those employed in Fig.~\ref{fig:comparison_direct_substitution}. The lattice results for the 24D and 32Dfine ensembles are juxtaposed with results obtained from dispersive analysis and perturbation theory. The lattice data below and above $\sqrt{Q^2}=0.2$ GeV are compiled using the substitution and direct methods, respectively.
         }
         \label{fig:m_qsq}
    \end{figure}

Fig.~\ref{fig:m_qsq} displays the lattice results for $M_n(Q^2)/\sqrt{Q^2}$ as a function of $\sqrt{Q^2}$, juxtaposed with findings from dispersive analysis~\cite{Seng:2018yzq,Seng:2018qru} and perturbation theory~\cite{Larin:1991tj,Baikov:2010je,Chetyrkin:2000yt}. 
Two curves from perturbation theory are presented. 
One is constructed using 4-flavor theory down to 1 GeV, while the other decouples the charm quark at 1.6 GeV and employs 3-flavor theory for the range $\mbox{1 GeV}\le
\sqrt{Q^2}\le \mbox{1.6 GeV}$. The observed discrepancy between these two curves suggests the unreliability of perturbation theory at small $\sqrt{Q^2}$.
Fortuitously, the contribution to $\Box_{\gamma W}^{VA}$ from this specific momentum range remains relatively small.
A substantially more significant contribution emerges from the $\sqrt{Q^2}<1$ GeV region, where the lattice results generally exhibit reduced uncertainties in comparison to those derived from dispersive analysis. At $\sqrt{Q^2}=0.2$ GeV, we compare the lattice results of $M_n(Q^2)/\sqrt{Q^2}$ obtained from the substitution and direct methods. For 24D and 32Dfine ensembles, the discrepancy is $0.006(49)$ GeV$^{-1}$ and $0.035(53)$ GeV$^{-1}$, indicating that the two approaches yield consistent outcomes. It's worth noting that these results are highly concordant with the findings presented in Table~\ref{tab:zero_mom}. This convergence is unsurprising, considering that the disparity between the two methods stems from the contrast between the lattice calculation of $\int d^3\vec{x}\,\bar{H}(t_g,\vec{x})$ and the PDG value of $-3g_A(\mu_p+\mu_n)$.

\subsection{Analysis of systematics}

In our calculation, the primary source of systematic effects arises from lattice discretization effects. 
We employ the lattice results of pion $\gamma W$ box contributions from our previous calculation~\cite{Feng:2020zdc} as inputs to estimate 
the residual lattice discretization effects after continuum extrapolation for the nucleon. 

\begin{figure}[htb]
     \centering
         \includegraphics[width=0.48\textwidth,angle=0]{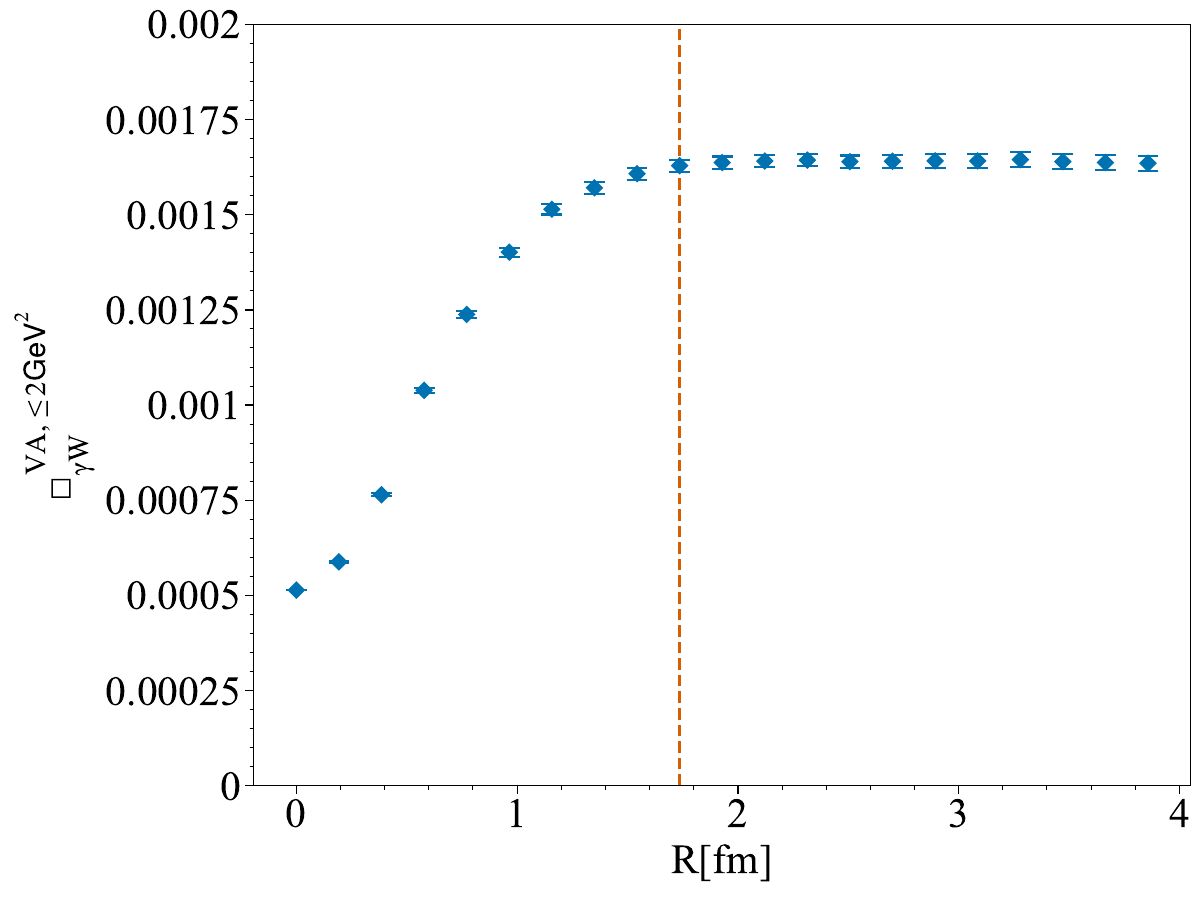}
         \caption{For ensemble 24D, lattice results of $\Box_{\gamma W}^{VA,\le 2\mathrm{GeV}^2}$ as a function of the integral range $R$. The plateau starts at
          $R\simeq 1.8$ fm, as indicated by the dashed line. Here, the substitution method is employed. Hence, $\Box_{\gamma W}^{VA,\le 2\mathrm{GeV}^2}$ is non-zero even at $R=0$ fm.
         }
         \label{fig:R_truncation}
    \end{figure}

To assess the finite-volume effect, we perform the spatial integral in $M_n^{\mathrm{SD}}(Q^2,t_s)$ and $M_n^{\mathrm{LD}}(Q^2,t_s,t_g)$ within a range of $|\vec{x}|\le R$.
Taking the ensemble 24D, which has better statistics, as an example, 
we depict $\Box_{\gamma W}^{VA,\le 2\mathrm{GeV}^2}$ as a function of $R$ in
Fig.~\ref{fig:R_truncation}. The final result is calculated at maximal $R\simeq 4$ fm, while the plateau is already evident at $R\simeq1.8$ fm, as indicated
by the dashed line in Fig.~\ref{fig:R_truncation}.
The deviation between $R=4$ fm and 1.8 fm is $\delta \Box_{\mathrm{FV}}=1.1\times10^{-5}$. We report this value as the finite-volume effect.

\begin{figure}[htb]
\centering
\includegraphics[width=0.48\textwidth,angle=0]{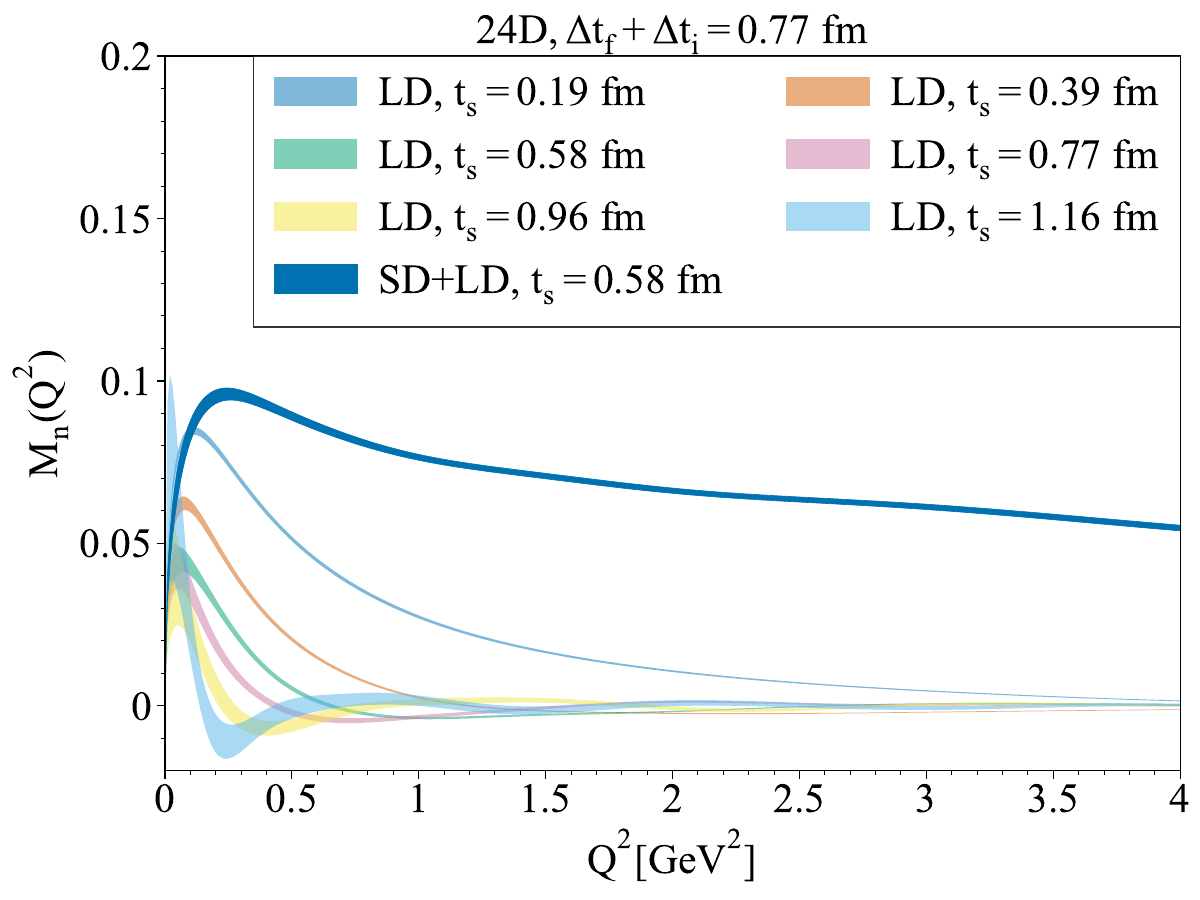}
\caption{$M_n^{\mathrm{LD}}(Q^2,t_s,t_g)$ as a function of $Q^2$ for ensemble 24D with various choices of $t_s$ but a fixed value of $t_g=0.58$ fm. Here, $\Delta t_i+\Delta t_f$ is set to 0.77 fm.
The error band denoted as ``SD+LD'' is the lattice result including both long-distance and short-distance contributions with $t_g=t_s=0.58$ fm.}
\label{fig:msd_qsq_LD}
\end{figure}

We employ the IVR method, reconstructing the long-distance contribution $M_n^{\mathrm{LD}}(Q^2,t_s,t_g)$ using the hadronic function $\mathcal{H}_{\mu\nu}^{VA}(t_g,\vec{x})$ as input.
Taking the ensemble 24D as an example, Fig.~\ref{fig:msd_qsq_LD} illustrates the results of $M_n^{\mathrm{LD}}(Q^2,t_s,t_g)$ as a function of $Q^2$ for different values of $t_s$ while keeping $t_g$ fixed at 0.58 fm. We observe that the smaller values of $t_s$ lead to larger the long-distance contribution and reduced statistical uncertainty. However, caution is needed not to choose excessively small values for $t_s$ or $t_g$, as either choice would potentially result in significant excited-state contamination.

\begin{figure}[htb]
\centering
\includegraphics[width=0.48\textwidth,angle=0]{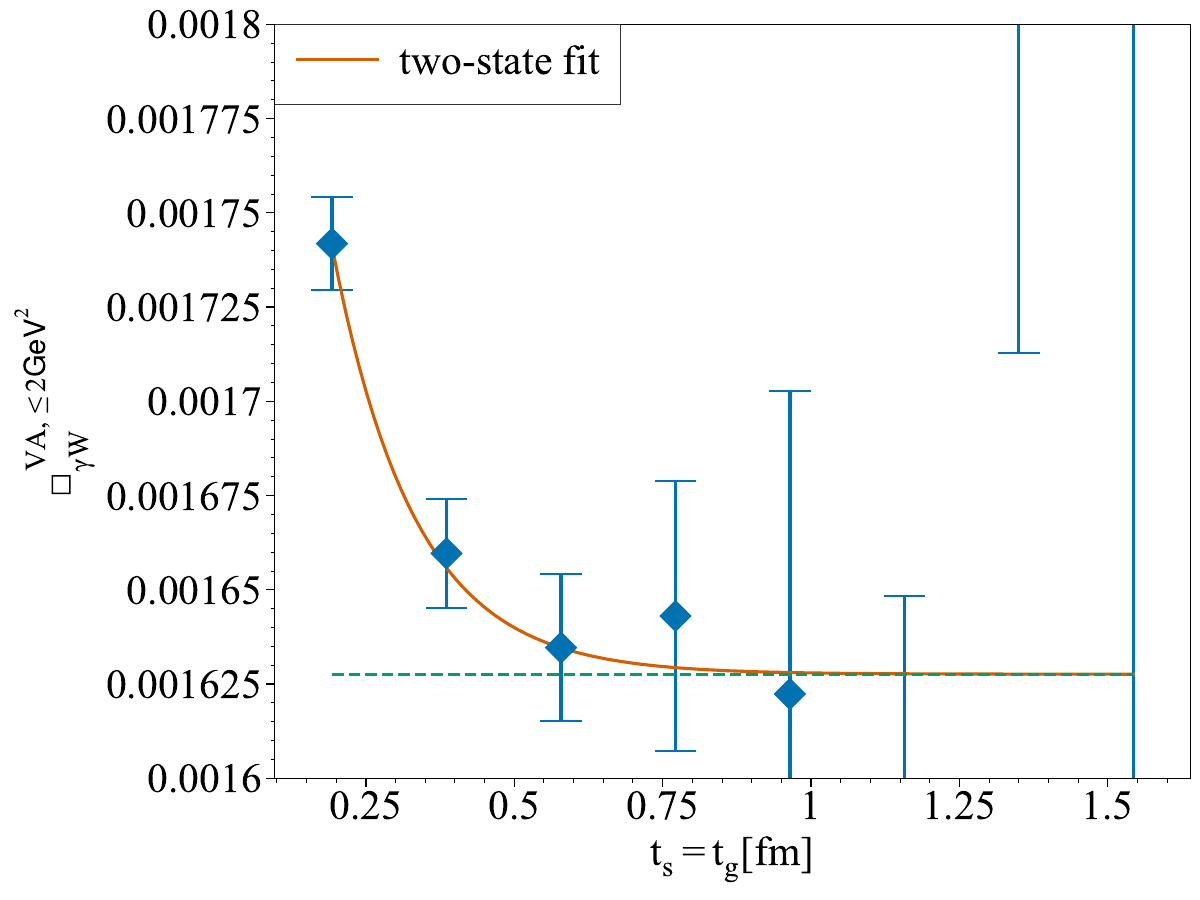}
\caption{$\Box_{\gamma W}^{VA,\le 2\mathrm{GeV}^2}$ as a function of $t_s=t_g$. We employ a two-state fit to examine the excited-state contaminations.}
\label{fig:t_dep}
\end{figure}

To assess the excited-state effects in the IVR approach, we make a plot of $\Box_{\gamma W}^{VA,\le 2\mathrm{GeV}^2}$ as a function of $t_s=t_g$ in Fig.~\ref{fig:t_dep}, once again using 24D as an example. We indeed observe the exited-state contamination at $t_s=t_g=0.19$ fm and 0.39 fm. The final result is calculated at $t_s=t_g=0.58$ fm, where the plateau has commenced. To estimate the residual excited-state effects, we
conduct a two-state fit. The discrepancy between the result at $t_s=t_g=0.58$ fm and the fit, $\delta \Box_{\mathrm{ES}}=1.2\times10^{-5}$, is reported  as the residual excited-state effect.

We have omitted the disconnected diagram, which vanishes in the flavor SU(3) limit. Comparable disconnected diagrams have been calculated in the $\pi\to\gamma^*\gamma^*$ form factors and their magnitude is approximately $1\%$ of that for connected diagrams~\cite{Gerardin:2023naa}. As a result, we attribute the systematic effects from disconnected diagrams to be $\delta \Box_{\mathrm{disc}}=1.5\times10^{-5}$.

\subsection{Continuum extrapolation for $\Box_{\gamma W}^{VV}-\Box_{\gamma W}^{VA}$}

\begin{figure}[htb]
\centering
\includegraphics[width=0.48\textwidth,angle=0]{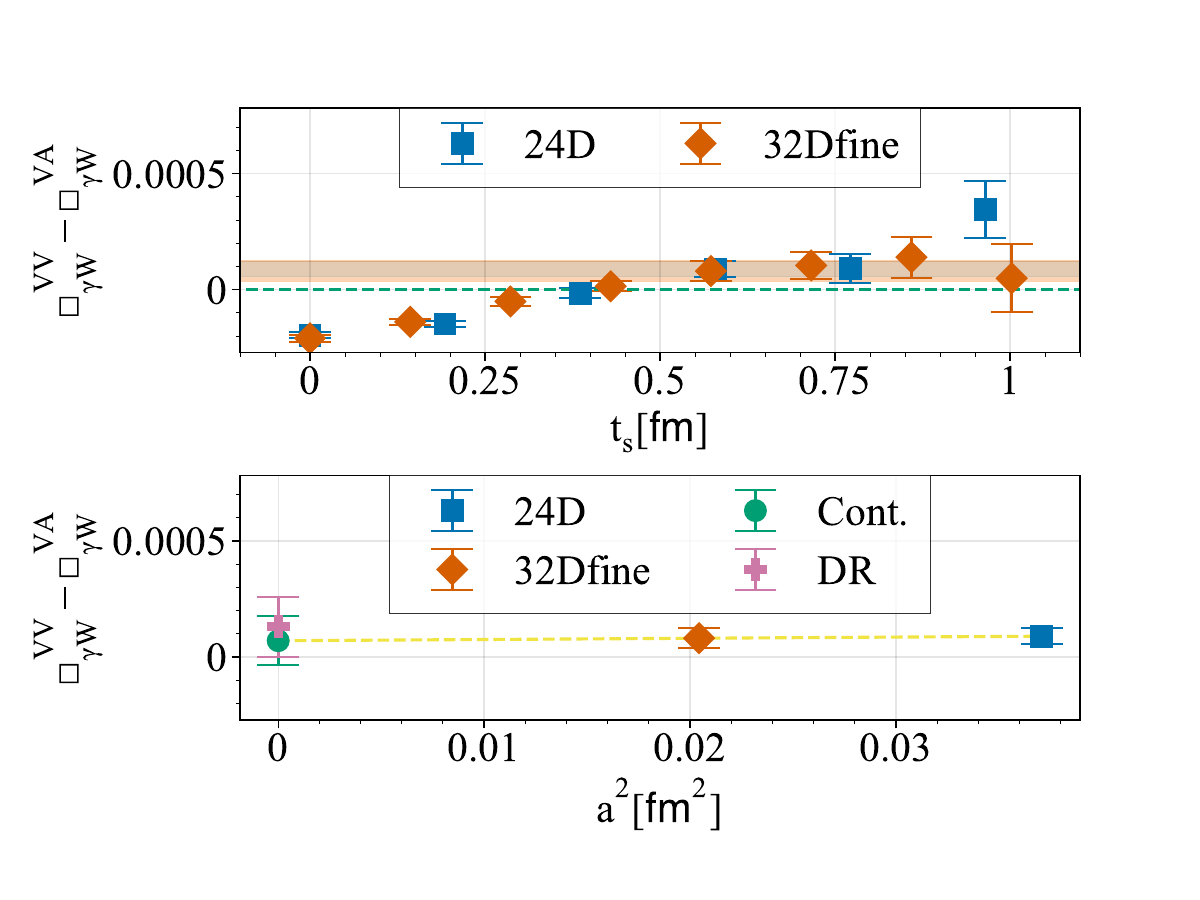}
\caption{In the upper panel, $\Box_{\gamma W}^{VV}-\Box_{\gamma W}^{VA}$ is depicted as a function of $t_s$. $\Delta t_i+\Delta t_f$ is set as $\sim0.75$ fm. In the lower panel, $\Box_{\gamma W}^{VV}-\Box_{\gamma W}^{VA}$ from the linear continuum extrapolation is juxtaposed with the outcome derived from dispersive analysis, denoted as ``DR'' here.
}
\label{fig:boxminus}
\end{figure}

In this study, we have calculated the $\gamma W$ box correction to $g_A$ using Eq.~(\ref{eq:corr_to_gA})~\cite{Hayen:2020cxh}.
The results are depicted in Fig.~\ref{fig:boxminus}.
Following a continuum extrapolation, we have determined $\Box_{\gamma W}^{VV}-\Box_{\gamma W}^{VA}=0.07(11)\times10^{-3}$, consistent with
the value of $0.13(13)\times10^{-3}$ derived from dispersive analysis~\cite{Gorchtein:2021fce}.

\end{document}